\documentclass[aip,jcp,reprint]{revtex4-1}
\usepackage{graphicx}
\usepackage{amsmath}
\usepackage{amssymb} 
\usepackage[usenames, dvipsnames]{color}

\begin{document}

\title{Simulations of ionic liquids confined by metal electrodes using periodic Green functions}

\author{Matheus Girotto}
\email{matheus.girotto@ufrgs.br}
\affiliation{Instituto de F\'isica, Universidade Federal do Rio Grande do Sul, Caixa Postal 15051, CEP 91501-970, Porto Alegre, RS, Brazil}

\author{Alexandre P. dos Santos}
\email{alexandre.pereira@ufrgs.br}
\affiliation{Instituto de F\'isica, Universidade Federal do Rio Grande do Sul, Caixa Postal 15051, CEP 91501-970, Porto Alegre, RS, Brazil}

\author{Yan Levin}
\email{levin@if.ufrgs.br}
\affiliation{Instituto de F\'isica, Universidade Federal do Rio Grande do Sul, Caixa Postal 15051, CEP 91501-970, Porto Alegre, RS, Brazil}

\begin{abstract}

We present an efficient method for simulating Coulomb systems confined by metal electrodes. 
The approach relies on Green functions techniques to obtain the electrostatic 
potential for an infinite periodically replicated system.  This avoids the use of image charges
or an explicit calculation of the induced surface charge, both of which dramatically slows down the simulations.
To demonstrate the utility of the new method we use it to obtain the ionic density profiles and the
differential capacitances, which are of 
great practical and theoretical interest, for a lattice model of an ionic liquid.

\end{abstract}

\maketitle

\newpage

\section{Introduction}
Simulations of Coulomb systems in confined geometries with a reduced symmetry are notoriously difficult.
This is due to the long range nature of the Coulomb interaction, which prevents the use of simple periodic boundary conditions. 
Instead, an infinite number of replicas must be considered, so that each particle in the simulation cell interacts with an infinite set of periodic replicas of itself and of all the other ions. To efficiently sum over the replicas the usual approach relies on Ewald summation techniques\cite{Ewa21,KoPe92,PeLe95,YoPe93,Fr02}.  Ewald methods have been implemented for both Coulomb and gravitational systems in $3$-d and various optimizations techniques have been developed. Unfortunately, when the symmetry of the system is reduced, which is the case when an interface is present, the computational cost of summation over replicas  increases dramatically due to the appearance of special functions and slow convergence\cite{Lek91,WiAd97,Maz05}. To overcome these problems a number of approaches have been proposed  \cite{DeCh98,Sp97,Kl92,KaMi01,ArDe02,YeBe99,GiLe16}.  The difficulty is that these methods are not easily generalized to systems bounded by metallic or dielectric surfaces.  The dielectric interfaces are important in many biophysics applications, while the 
metallic electrodes are omnipresent in electrochemistry and play a fundamental role in the discussion of ionic liquids which, due to their use in renewable energy storage devices, are of great practical and technological importance\cite{Wi09,HeYa14,SiGo08,SiGo10}.  
To simulate metallic surfaces of electrodes various approaches have been proposed\cite{RoRo13,SiSp95,LaMa07}. Unfortunately, all are computationally very expansive, relying on a minimization procedure to calculate the induced surface charge at every molecular dynamics time step\cite{ChRo14,FeKo14}. In the present paper we propose a completely new method for studying ionic systems confined by parallel metal surfaces. The formalism is based on the periodic Green functions which are constructed to satisfy the appropriate Dirichlet boundary conditions. The resulting infinite sum over replicas is fast convergent and can be truncated at a reasonably small number of replicas. To show the utility of the method, we apply it to calculate the ionic density profiles and the differential capacitances  of a lattice model of a room temperature ionic liquid, confined by two metal electrodes.
\begin{figure}[t]
\begin{center}
\includegraphics[scale=0.5]{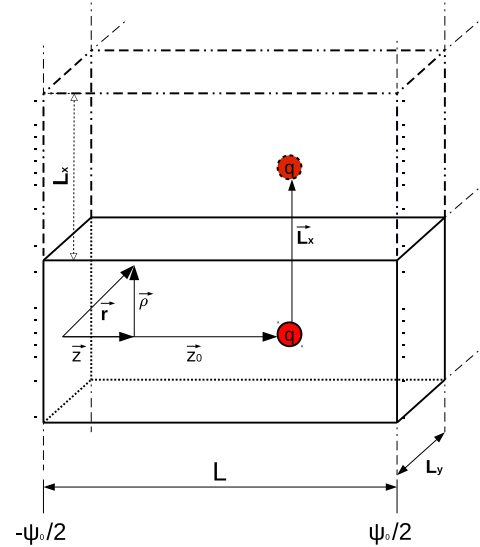}
\end{center}
\caption{Point charge $q$ located at $z_0 \hat{{\pmb z}}$ between two infinite metal electrodes with constant potential difference $\psi_0$. The electrodes are at $z=0$ and $z=L$. The electrostatic potential is calculated at a point ${\pmb r}$ indicated in cylindrical coordinates. The dashed lines show the first replica of the simulation box in the $x$ direction.}
\label{fig1}
\end{figure}

\section{The Theory}

Our goal is to calculate the electrostatic potential inside a simulation cell of 
size $L_x \times L_y \times L$, bounded by conducting surfaces separated by distance $L$.
For an ion located at $(x_0,y_0,z_0)$ inside the main simulation cell, there will be an infinite set of replicas located at $(x_0 \pm m_x L_x,y_0 \pm m_y L_y,z_0)$, with $m_x,m_y \in  \mathbb{Z}^+$, see Fig \ref{fig1}.

We start by calculating the electrostatic potential at position ${\pmb r}=\rho\hat{{\pmb \rho}}+\varphi\hat{{\pmb \varphi}}+z\hat{{\pmb z}}$ produced by a \textit{single} ion of charge $q$ located at ${\pmb r'}=z_0\hat{{\pmb z}}$, see Fig \ref{fig1}, between two parallel grounded
infinite metal electrodes. The final solution will be easily generalized to an arbitrary potential difference between the boundaries. We use cylindrical coordinates in order to explore the azymuthal symmetry of the potential, eliminating the $\varphi$ dependence of it.
To obtain the electrostatic potential requires 
us to solve the Poisson equation\cite{Jac99}
\begin{equation}
\nabla ^2 \phi({\pmb r},{\pmb r'}) = - \frac{4\pi q}{\epsilon}\delta({\pmb r} -{\pmb r'}) \ ,
\label{eqlaplace}
\end{equation}
with the  Dirichlet boundary condition $\psi_0=0$ at each surface. 
We start by expanding the delta function in the eigenfunctions of the differential operator 
\begin{equation}
 \frac{d^2 \psi_n}{d z^2}+k_n^2 \psi_n=0 \ ,
\label{dif}
\end{equation}
satisfying the boundary conditions $\psi_n(0)=\psi_n(L)=0$.  
The eigenfunctions are found to be $\psi_n(z)=\sqrt{2/L} \sin (k_n z)$, with $k_n= n \pi/L$. 
The delta function can then be 
written as
\begin{equation}
\delta(z-z_0)=\frac{2}{L}\sum_{n=1}^{\infty}\sin (\frac{n\pi z}{L})\sin (\frac{n\pi z_0}{L}) \,.
\label{delta1}
\end{equation}
The electrostatic potential can now be written as
\begin{equation}
\phi(\rho,z;z_0)=\frac{2q}{\epsilon L}\sum_{n=1}^{\infty}\sin (\frac{n\pi z}{L})\sin (\frac{n\pi z_0}{L})g_{n}(\rho) \,.
\label{green1}
\end{equation}
Substituting this expression into Eq.~\ref{eqlaplace} we obtain an ordinary differential equation for $g_{n}(\rho)$,
\begin{equation}
\frac{1}{\rho}\frac{\textrm{d}}{\textrm{d}\rho}(\rho\frac{\textrm{d}g_{n}}{\textrm{d}\rho})-k_n^2 g_{n}=-\frac{2}{\rho}\delta(\rho) \,,
\label{bessel}
\end{equation}
which has modified Bessel functions of order zero as solutions, $g_{n}(\rho)=AI_0(k_n\rho)+BK_0(k_n\rho)$.  Since the potential must vanish as $\rho \rightarrow \infty$, the coefficient $A=0$, while the coefficient $B$ is determined by the singular part of the potential, and is found to be $B=2$. The electrostatic potential produced by an ion located at ${\pmb r}=z_0\hat{{\pmb z}}$ between two grounded  metal surfaces is then\cite{Jac99}
\begin{equation}
\phi(\rho,z;z_0)=\frac{4q}{\epsilon L}\sum_{n=1}^{\infty}\sin(k_nz)\sin(k_nz_0)K_0(k_n\rho) \ .
\label{potsum}
\end{equation}
For simulating an ionic system we will need to periodically replicate the main simulation cell.  Since the electrostatic potential in Eq. (\ref{potsum}) satisfies the appropriate boundary conditions, 
the electrostatic potential produced by an ion located at $(x_0,y_0,z_0)$ and all of its periodic replicas can be simply obtained by a superposition.  We find,
\begin{equation}
\begin{split}
&G({\pmb r};{\pmb r}_0)=\frac{4q}{\epsilon L}\sum_{{\pmb m}=-\infty}^\infty\sum_{n=1}^{\infty}\sin(k_n z)\sin(k_nz_0) \times \\
&K_0\left(k_n\sqrt{(x-x_0 + m_x L_x)^2+(y-y_0 + m_y L_y)^2}\right) \ .
\end{split}
\label{potsumtrans}
\end{equation} 
The modified Bessel function $K_0(x)$ decays exponentially for large $x$, therefore, in practice we will need only a small number of replicas to obtain the electrostatic potential to any desired accuracy.  Unfortunately, Eq. (\ref{potsumtrans}) is ill defined when $x=x_0$ and $y=y_0$. The problem arises because $K_0(x)$ has a logarithmic divergence at $x=0$, which manifests itself in $m_x=m_y=0$ term of Eq. (\ref{potsumtrans}). This term corresponds to the electrostatic  potential arising from the ion inside the main cell. Hence, the limiting value as $\rho\rightarrow 0$ of $G({\pmb r};{\pmb r}_0)$ is complex to obtain in this Green function representation. To overcome this difficulty we will use a different representation of the Green function to calculate the electrostatic potential produced by this ion.  

Once again we consider {\it one} (no replicas) ion located between two infinite grounded conducting surfaces at ${\pmb r}=z_0\hat{{\pmb z}}$.  We now use the following representation of the delta function
\begin{equation}
\frac{1}{\rho}\delta(\rho) = \int_0^{\infty}kJ_0(k\rho)\textrm{d}k \ ,
\label{delta2}
\end{equation}
where $J_0$ is the Bessel function of order zero.  The electrostatic potential can now be written as
\begin{equation}
\phi(\rho,z;z_0)= \frac{q}{\epsilon}\int_0^{\infty}k J_0(k\rho) g_k(z,z_0) \textrm{d}k \ .
\label{green2}
\end{equation}
Substituting Eq.~\ref{green2} into Eq.~\ref{eqlaplace},
we obtain an ordinary differential equation for $g_k(z,z_0)$:
\begin{equation}
\frac{\textrm{d}^2g_k}{\textrm{d}z^2}-k^2g_k=-2 \delta(z-z_0) \ .
\label{eqexp}
\end{equation}
Applying the boundary conditions, we finally obtain
\begin{equation}
\begin{split}
&\phi(\rho,z;z_0) = \frac{q}{\epsilon}\int \textrm{d}kJ_0(k\rho) \times \\
&\frac{e^{k|z-z_0|-2kL}+e^{-k|z-z_0|}-e^{-k(z+z_0)}-e^{k(z+z_0)-2kL}}{1-e^{-2kL}} \,,
\end{split}
\label{potint}
\end{equation}
which is well behaved when $\rho\rightarrow 0$, as long as $z \ne z_0$. 
This expression is equivalent to Eq. (\ref{green1}) and can be used to calculate the electrostatic 
potential produced by the ion inside the main simulation cell, replacing 
the $m_x=m_y=0$ term of Eq. (\ref{potsumtrans}), since it will rapidly converge even if the potential is to be calculated at $\rho=0$.

Suppose an ion is placed at $z=z_0$ between two 
infinite grounded metal surfaces.  How much charge will be induced on each electrode?
The surface charge density on the left electrode is
\begin{equation}
\sigma(\rho) = -\frac{\epsilon}{4\pi}\frac{\partial \phi}{\partial z}\Biggr|_{z=0}=-\frac{q}{2\pi}\int_0^{\infty}\textrm{d}kJ_0(k\rho)k\frac{\sinh[k(L-z_0)]}{\sinh(kL)} \ ,
\end{equation}
and the total charge is 
\begin{equation}
Q_l^0= -\frac{q}{2\pi}\int \textrm{d}\varphi \int \rho \textrm{d}\rho \int \textrm{d}k k J_0(k\rho) \frac{\sinh[k(L-z_0)]}{\sinh(kL)} \ .
\label{pint}
\end{equation}
Eq. (\ref{pint}) is conditionally convergent.  To conveniently perform the integral we introduce  a convergence factor $e^{-\alpha \rho}$  which allows us to change the order of integration. Performing the integration first over $\rho$, and then changing variables and taking the $\alpha\rightarrow 0$ limit, we find that the total charge on the left electrode is
\begin{equation}
Q_l^0= - q (1 - \frac{z_0}{L}) \ .
\label{qplate}
\end{equation}
Similarly the surface charge on the right electrode is $Q_r^0=-q z_0/L$. 
\begin{figure}[t]
\begin{center}
\includegraphics[scale=0.325]{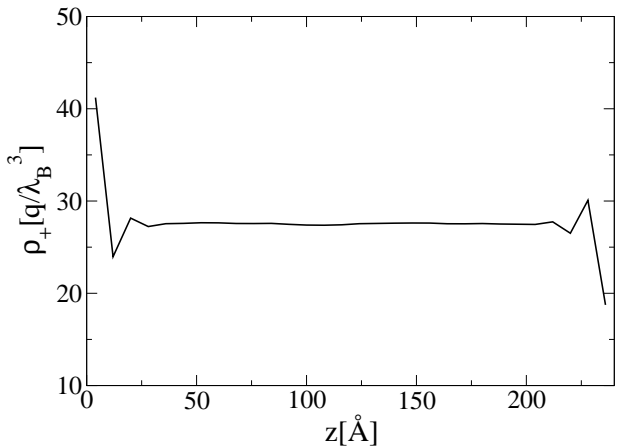}
\end{center}
\caption{Cationic profile of ionic liquid between the electrodes. The parameters are: $\gamma=\frac{1}{2}$, $\lambda_B=38.4$\AA$\,$ and $\sigma=0.05V$.}
\label{fig2}
\end{figure}

So far our discussion has been restricted to the grounded metal surfaces.  Often, however,
the electrostatic potential difference between the electrodes is controlled by an external battery, so that the potential of the electrode located at $z=0$ is fixed at $-\psi_0/2$ and of the electrode located at $z=L$ at $+\psi_0/2$.  Using the uniqueness property of the 
Laplace equation, it is simple to account for the surface potential controlled by an external source. We observe that if we add to Eq. (\ref{potsumtrans}) a potential
\begin{equation}
\phi_s(z)=\left(\frac{z}{L}-\frac{1}{2}\right)\psi_0 \,,
\label{sufpot}
\end{equation} 
the sum will satisfy the Laplace equation with the appropriate boundary conditions, providing a unique solution. For a periodically replicated charge neutral system with $N$ ions at positions $\{{\pmb r}_i \}$ and  electrodes
held at potentials $\mp \psi_0/2$, respectively, the total charge 
on the left and right electrodes {\it within the simulation cell} will then be
\begin{equation}
Q_{l,r}= \mp\frac{\epsilon \psi_0 A}{4 \pi L}\pm\sum_{i=1}^Nq_i\frac{z_i}{L} \,,
\label{char}
\end{equation} 
where $A=L_xL_y$ is the area of the electrode inside the simulation cell. Note that $Q_l=-Q_r$.

\section{Simulations}

We are now in a position to perform simulations of N-body Coulomb systems confined by 
two parallel metal electrodes.The object of particular interest for the ionic liquids community is the differential capacitance, which can be obtained from the fluctuations of the surface charge on the electrodes\cite{RoRo13}.  
The partition function
in the fixed electrostatic potential ensemble is 
\begin{equation}
\mathcal{Z}_{\psi}=\int \prod_{i=1}^N \textrm{d}{\pmb r_i} \int \textrm{d} Q e^{-\beta[E({\pmb r_1},...,{\pmb r_N},Q)  - \psi Q]} \ ,
\label{partfunc}
\end{equation} 
where $\beta=1/k_B T$ and the surface charge on the left and right electrodes is $\mp Q$, respectively. 
Note that in this ensemble the surface charge on the electrodes is allowed to fluctuate.  
The differential capacitance of the system can then be calculated straightforwardly as
\begin{equation}
C =\frac{1}{A} \frac{\partial \left<Q\right>}{\partial \psi}=\frac{1}{\beta A}\Big(\frac{\partial^2 \ln \mathcal{Z}_{\psi}}{\partial \psi^2}\Big) =  \frac{\beta}{A} [\left<Q^2\right>-\left<Q\right>^2] \ .  
\label{DC}
\end{equation}
It is important to note that in order to perform a simulation at a {\it fixed electrostatic
potential}, we need to know the total electrostatic energy $E(Q)$ of a system  with electrodes 
carrying a {\it fixed amount} of surface charge $-Q$ and $+Q$, respectively. Since the electrodes are metallic, they must be equipotential.  This means that the distribution of the surface charge 
will {\it not} be uniform and will respond to ionic motion.  
For a given $Q$, the surface potential $\psi_0$ will, therefore, 
fluctuate. Since the system is charge neutral, the surface potential for a given 
ionic distribution inside the simulation cell can be easily calculated using Eq. (\ref{char}),
\begin{equation}
\psi_0 = \frac{4 \pi L}{\epsilon A}\left(Q + \sum_{i=1}^Nq_i\frac{z_i}{L}\right).  
\label{psiup}
\end{equation}
The total electrostatic energy inside the simulation cell is then
\begin{equation}
E(Q)= \frac{1}{2}\sum_{i \ne j}^N q_i G({\pmb r_i};{\pmb r_j})+
\sum_{i=1}^N \left[ U_s({\pmb r_i}) + \frac{1}{2} q_i \phi_s(z_i) \right] + \frac{1}{2} \psi_0 Q,  
\label{totalE}
\end{equation}
where the periodic Green function is given by Eq. (\ref{potsumtrans}) with $m_x=m_y=0$ term 
replaced by Eq. (\ref{potint}), and the self energy of an ion at ${\pmb r_i}$ is
\begin{equation}
U_s({\pmb r_i}) = \frac{q_i}{2}\lim_{\rho\to 0} \left[G({\pmb r_i};{\pmb r_i}) - \frac{q_i}{\epsilon \rho}\right] \ .
\label{lim}
\end{equation}
Using the identity
\begin{equation}
\int_0^\infty \textrm{d}kJ_0(k\rho)= \frac{1}{\rho}\,,
\end{equation}
the limit in Eq.(\ref{lim}) can be performed explicitly \cite{Lev06}, resulting in
\begin{equation}
\begin{split}
& U_s({\pmb r_i}) = \frac{q^2}{2\epsilon}\int \textrm{d}k
\frac{2 e^{-2kL}-e^{-2 k z_i}-e^{2 k z_i-2kL}}{1-e^{-2kL}}+\\
&\frac{2q^2}{\epsilon L}\sum_{{\pmb m} \ne {\pmb 0}}^\infty \sum_{n=1}^{\infty}\sin^2(k_nz_i) K_0\left(k_n\sqrt{m_x^2 L_x^2+ m_y^2 L_y^2}\right) \ .
\end{split}
\label{us}
\end{equation}

In practice since $K_0(x)$ decays exponentially for large $x$, the sums 
in Eq.(\ref{lim}) converge very fast.  To demonstrate the utility of the present
method we study a Coulomb lattice gas\cite{KoFi02,RaSh93,DaBi15}  confined  between
two electrodes held at potential $\mp \psi/2$, respectively. The Monte Carlo simulations are performed using the Metropolis algorithm.  To further speed up the simulations we have pre-calculated the electrostatic potentials at each lattice position at the beginning of the simulation. 
We allow swap moves between the ions and between the ions and the empty sites, during which the surface
charge on the electrode remains fixed and the energy of the system $E(Q)$ is calculated 
using Eq.(\ref{totalE}).   We also allow 
moves in which the surface charge on the electrodes increases or decreases in accordance with the Boltzmann factor of Eq.(\ref{partfunc}).
The simulations are performed in a cell of volume $V=L_xL_yL$, with $L_x=L_y=80\AA$ and $L=3L_x$. The lattice gas is confined in the region $-L_x/2<x<L_x/2$, $-L_y/2<y<L_y/2$ and $0<z<L$. The negatively charged electrode is positioned at $z=0$ and the positive one at $z=L$. We define the Bjerrum length as $\lambda_B=q^2/k_BT\epsilon$, and consider two specific values $\lambda_B=7.2$\AA$\,$ and $\lambda_B=38.4$\AA. The first value is appropriate for room temperature electrolytes while the the second is for room temperature ionic liquids \cite{WaWe03,Ko08,DrWe11}, which have  dielectric constant around $\epsilon=15$.  The concentration of ionic liquid is controlled by the compacity factor $\gamma=(N_++N_-)/(N_++N_-+N_0)$, where $N_+$ is the number of cations, $N_-$ the number of anions, and $N_0$ the number of voids. We will set $\gamma$ to $\frac{1}{20}$ for electrolytes, and $\frac{1}{2}$ for ionic liquids. The lattice spacing is set to $8$\AA, characteristic of ionic diameter. For now we consider a symmetric case with charge of cation $q$ and charge of anions $-q$, where $q$ is the charge of the proton. The model, however, can be easily extended to asymmetric ionic liquids. In the simulations we have used around $\approx 10^4$ $ {\pmb m}$-vectors in the energy computation. The averages were calculated with $5\times 10^4$ uncorrelated samples after equilibrium was achieved.  
\begin{figure}[t]
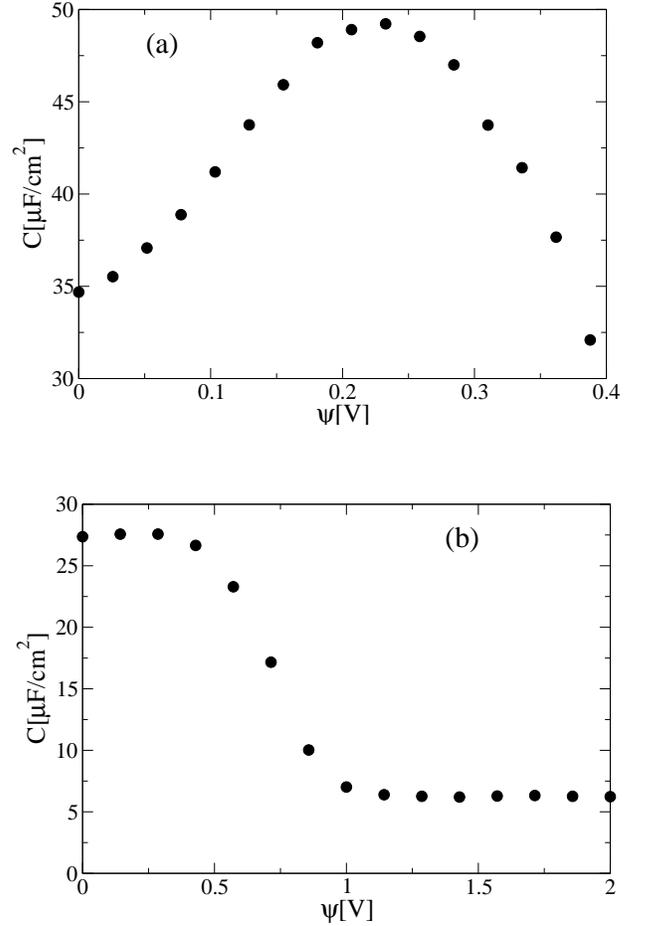

\begin{center}
\includegraphics[scale=0.325]{fig3a.eps}\vspace{0.85cm}

\includegraphics[scale=0.325]{fig3b.eps}%
\end{center}
\caption{Differential capacitance  calculated using Eq.~\ref{DC}. Panel (a) shows the electrolyte regime  with parameters $\gamma=\frac{1}{20}$ and $\lambda_B=7.2$\AA; and (b) shows the typical bell-shaped differential capacitance of ionic liquids, $\gamma=\frac{1}{2}$ and $\lambda_B=38.4$\AA.}
\label{fig3}
\end{figure}

In Fig.~\ref{fig2} we show the oscillatory behavior of the counterion density profile near an electrode\cite{FeKo14}, and in Fig.~\ref{fig3} we present 
the differential capacitance  in electrolyte and ionic liquid regimes. 
Fig.~\ref{fig3} (a) shows the characteristic minimum of differential capacitance at zero potential predicted by the Poisson-Boltzmann theory, followed by a maximum for higher applied voltages. The behavior is characteristic of electrolyte solutions \cite{MaBa62}. On the other hand in the regime of ionic liquids, where steric and electrostatic correlations play the dominant role \cite{Lev02}, the behavior is quite different~\cite{Kor07,KoHo14,KOHo142}.  Fig.~\ref{fig3} (b) shows that unlike electrolytes, ionic liquids have a maximum of differential capacitance at $\psi=0$V. It is gratifying to see that
a simple lattice model captures this complicated transition of differential capacitance between electrolyte and ionic liquid regimes.  

\section{Conclusions}

We have presented a new method for simulating ionic systems confined by infinite metal electrodes. Our algorithm is based on periodic Green functions derived in the present Letter.  The main advantage of the method is that it avoids the explicit calculation of the potential produced by the infinite distribution of the image charges or a numerical calculation of the induced surface charge at each simulation time step.  Furthermore, since 
the potential produced by the ions is effectively screened by the electrodes, we only need a small number of replicas to achieve any desired precision.  As a demonstration of the utility of the method, we have applied it to the calculations of the differential capacitance of a Coulomb gas, both in the electrolyte and ionic liquid regimes.  In the future work we will extend these calculations to continuum ionic liquids.

\section{Acknowledgements}

This work was partially supported by the CNPq, INCT-FCx, and by the US-AFOSR under the grant 
FA9550-16-1-0280.

\section{References}

\bibliography{ref}

\end{document}